\documentclass[%
amsmath, amssymb, aps, prb, floatfix, longbibliography, twocolumn
]{revtex4-2}

\usepackage{MnSymbol}
\usepackage{amsmath,graphicx,color}
\usepackage{natbib}
\usepackage{placeins}
\setcitestyle{square}
\usepackage{braket}
\usepackage{makerobust}
\usepackage{xcolor}
\usepackage{hyperref}
\hypersetup{%
 colorlinks,
 breaklinks=true,
 plainpages=false,%
 citecolor=blue,
 linkcolor=blue,
 urlcolor=blue,
 bookmarksopen=true,%
 bookmarksnumbered=false,%
 bookmarksdepth=5%
}

\newcommand{\makeauthor}[2]{\newcommand{#1}[1]{{%
 \sffamily\color{#2}{%
 \bfseries\begingroup\escapechar=-1\edef\x{\endgroup\string#1}\x:%
 } ##1}}%
 \MakeRobustCommand#1}
\makeauthor{\rew}{blue}
\makeauthor{\jb}{purple}
\makeauthor{\nk}{red}

\begin{document}

\title{\textbf{Topological Altermagnetic Insulators}}

\author{Jasmin Bedow, Nitin Kaushal and Marcel Franz}
\affiliation{Department of Physics and Astronomy and Quantum Matter Institute, University of British Columbia, Vancouver, British Columbia BC V6T 1Z4, Canada}

\date{\today}

\begin{abstract}
We study the emergence of altermagnetic topological phases stabilized by Ising spin-orbit coupling in a prototypical Lieb lattice model of a correlated altermagnet. 
Treating the electron interactions within the Hartree-Fock and the exact diagonalization techniques, we establish the co-existence of a quantum spin Hall effect with altermagnetic spin order over a wide region of the parameter space for average electron densities of 2 and 4 per unit cell. We explore how the magnetic structure along edges influences the electronic behavior of the associated topological edge modes, demonstrating their robustness against inversion-symmetry breaking terms. 
\end{abstract}

\maketitle

{\it Introduction.} The combination of zero net magnetization and non-relativistic spin splitting in the electronic bands of altermagnets  \cite{Hayami2019,Smejkal2020,Yuan2020,Mazin2021,Smejkal2022,Smejkal2022_2} makes them an exciting platform for the creation of topological phases in two and three dimensions \cite{Zhu2023,Heung2024,Ghorashi2024,Hodge2025,Li2025,Chen2025,Feng2025,Antonenko2025}. Adding the Ising-type spin-orbit coupling (SOC) to minimal non-interacting models can realize quantum spin Hall effects \cite{Zhang2025,Antonenko2025,Chen2025} as well as corner modes associated with higher-order topology \cite{Yang2025}, making them fascinating candidate platforms for applications in spintronics \cite{Bai2023,Ang2023,Sun2023,Chi2024,Zhang2024} and quantum computing \cite{Heung2024,Hodge2025}. In particular, Ref. \cite{Antonenko2025} showed that the interplay between two-dimensional altermagnetic semimetals with an out-of-plane N\'eel vector and an Ising-type spin-orbit coupling gives rise to a topological altermagnetic insulator (TAI), where altermagnetic order co-exists with a non-trivial spin Chern number.

Yet, an experimental realization of this topological altermagnetic phenomenon remains outstanding despite there being many candidate materials for altermagnets  \cite{Krempasky2024,Lee2024,Osumi2024,Mazin2023,Ding2024,Lu2025,Li2025,Ni2010,Fuwa2010,Freelon2019,Wei2025,Jiang2025,Zhang2025}. The key observations include $g$-wave altermagnetic band splitting in MnTe \cite{Krempasky2024,Lee2024,Osumi2024} and CrSb \cite{Ding2024,Lu2025,Li2025} and $d$-wave altermagnetic band splitting in the group of the oxycalcogenides  \cite{Ni2010,Fuwa2010,Freelon2019,Wei2025,Jiang2025,Zhang2025}. In the latter family, it was recently observed in neutron diffraction experiments that certain vanadate compounds exhibit G-type antiferromagnetism \cite{Sun2025_2,Xie2026}, which would enforce spin-degenerate bands. The combined neutron diffraction and angle-resolved photoemission spectroscopy (ARPES) studies \cite{Yang2026}, can be naturally interpreted as bulk antiferromagnets hosting surface altermagnetism \cite{Lange2026}. 
The central ingredient for the surface altermagnetism lies in the ``anti-CuO$_2$" structure of their V$_2$O layers in a Lieb lattice arrangement, suggesting that the altermagnetic spin splitting would persist in the monolayer limit.
For such monolayers, it was recently shown theoretically that a Lieb lattice Hubbard model for materials with this crystallographic arrangement can realize both altermagnetic Mott insulators and metals as its ground state \cite{Kaushal2025}. 

In this Article, we show that this prototypical model for 2D altermagnets together with an Ising-type SOC can realize the proposed topological altermagnetic insulating state in a large region of the parameter space. 
Therein, we do not assume the altermagnetic order, but rather obtain the magnetic ordering self-consistently from the interactions using Hartree-Fock (HF) theory, and demonstrate that the Ising SOC leads to a preferred direction of the N\'eel vector. We show that topology and altermagnetism can emerge simultaneously from a minimal model, paving the way towards finding a suitable material candidate. 
For this purpose, we study the Lieb lattice Hubbard model with local, repulsive Hubbard interactions present on the $B$ and $C$ sublattice, containing the magnetic atoms. We keep the $A$ sublattice non-magnetic and include an Ising-type SOC between the $B$ and $C$ sublattices. Using the HF decomposition, we then map out the topological and magnetic phase diagram as a function of the local Hubbard interaction for electron densities of 2 and 4 per unit cell. 
We study the electronic structure for different edge terminations to assess the spectroscopic signatures of the TAI and investigate its robustness against Rashba SOC along different bonds. Finally, we validate our mean-field results using exact diagonalization (ED) techniques.

\begin{figure}
    \centering
    \includegraphics[width=0.95\linewidth]{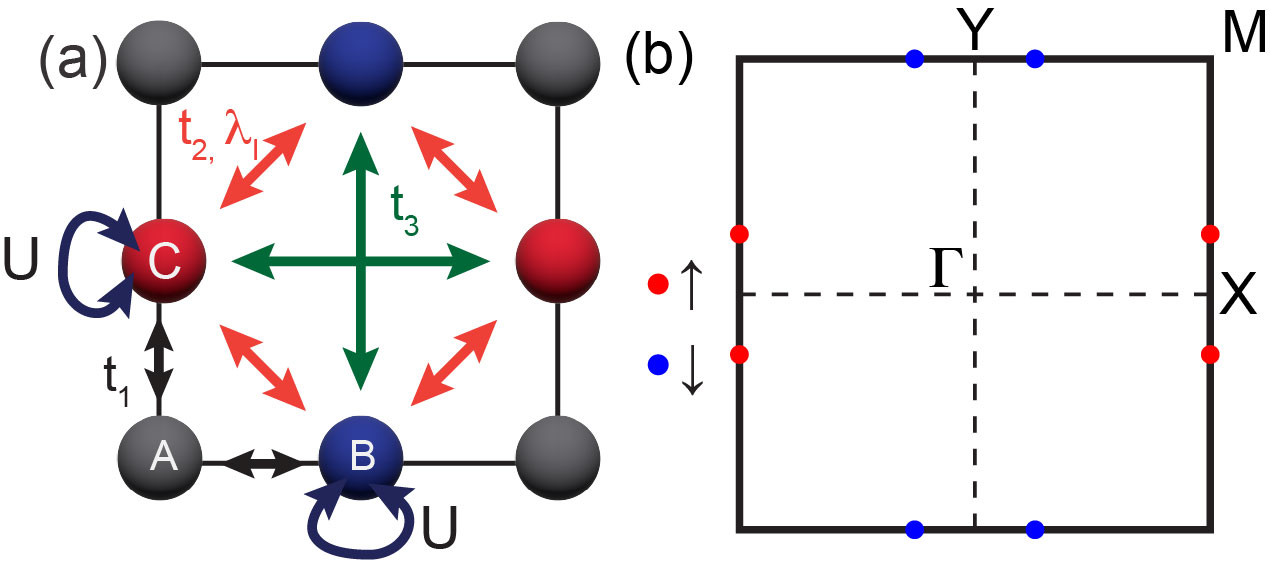}
    \caption{(a) Schematic of the Lieb lattice Hubbard model used in this study, and (b) the Brillouin zone with the locations of the gapped Dirac points in the TAI phase marked.}
    \label{fig:Fig0}
\end{figure}

{\it Model.} We investigate the two-dimensional Lieb-lattice Hubbard model, described by the Hamiltonian
\begin{equation}
\begin{aligned}
    \mathcal{H}_0 =& \;  \varepsilon_A \sum_{i \in A} n_{i} -\mu \sum_{i} n_i + U \sum_{i \in \{B,C\}} n_{i,\uparrow} n_{i,\downarrow}  \\
    &+ t_1 \sum_{\langle i,j \rangle, \alpha} c^\dagger_{i,\alpha} c^{}_{j,\alpha} + \sum_{\langle\langle i,j \rangle\rangle, \alpha} t_2 c^\dagger_{i,\alpha} c^{}_{j,\alpha} \\
    &+ t_3 \left( \sum_{i \in B ,\alpha} c^\dagger_{i,\alpha} c^{}_{i+\hat{y},\alpha} + \sum_{i \in C,\alpha} c^\dagger_{i,\alpha} c^{}_{i+\hat{x},\alpha} + h.c.\right)
    \label{eq:ham_liebHubbard}
\end{aligned}
\; ,
\end{equation}
which we show schematically in Fig.~\ref{fig:Fig0} (a) with the sublattices $A,B,C$ labelled.

Here, $c^\dagger_{i,\alpha}$ creates an electron at site $i$ (labelling both the unit cell and the sublattice) of spin $\alpha$, $n_{i,\alpha} = c^\dagger_{i,\alpha} c^{}_{i,\alpha}$ is the density operator and $n_i = \sum_\alpha n_{i,\alpha}$. $\mu$ is the chemical potential, which we determine such that the desired filling is achieved, $\varepsilon_A$ is a local potential at site $A$. $t_1$, $t_2$ and $t_3$ describe the amplitude of the first, second and third neighbor hopping parameters, while $U$ denotes the strength of the Hubbard repulsion.

For $t_3 = 0$, it was previously shown \cite{Kaushal2025} using unrestricted HF theory at filling 2 and 4, that the interplay of $t_1$ and $t_2$ with the Hubbard interaction $U$ can lead to antiferromagnetic order on sublattices $B$ and $C$, either as a metal or an insulator. As a result, the system becomes a crystallographic altermagnet with the effective time-reversal symmetry $C_{4z} \mathcal{T}$, where $C_{4z}$ denotes the fourfold rotation around the $z$-axis and $\mathcal{T}$ is the usual time-reversal operator.

{\it Topological Altermagnetic Insulators.} For a topological phase to emerge, we consider the effects of an Ising-type SOC between the $B$ and $C$ sublattices
\begin{equation}
    \mathcal{H}_{\lambda_I} = \mathrm{i} \lambda_{I} \sum_{\langle\langle i,j \rangle\rangle,\alpha} (-1)^\alpha  \nu_{ij} c^\dagger_{B,i,\alpha} c_{C,j,\alpha} + h.c.,
\end{equation}
where, $\nu_{ij} = \pm 1$, chosen such that the system has $C_{4z} \mathcal{T}$ symmetry.
It was previously shown \cite{Antonenko2025} that this can lead to an altermagnetic Dirac semimetal on the Lieb lattice becoming a topological insulator with a non-trivial spin Chern number $C_s = C_\uparrow - C_\downarrow$ with \cite{Prodan2009}
\begin{equation}
   C_\sigma = \frac{1}{2\pi \mathrm{i}} \int_{\mathrm{BZ}} \mathrm{d}^2 \mathbf{k} \; \mathrm{Tr} \left\{ P_\sigma (\mathbf{k}) \left[ \partial_{k_x} P_\sigma (\mathbf{k}),  \partial_{k_y} P_\sigma (\mathbf{k}) \right] \right\} \; ,
\end{equation}
where $P_\sigma(\mathbf{k})$ refers to the projector onto the occupied bands with spin $\sigma$. 
TAI emerges when the N\'eel vector points out of plane, which opens a topological gap at the Dirac points.

In the previous work \cite{Antonenko2025} the N\'eel vector direction was simply chosen to be out-of-plane. Here we determine it from our interacting model in an unbiased way. To this end  
 we evaluate the staggered magnetization in the $xy$-plane and along the $z$-axis defined as
\begin{align}
    m_{s,z} &= |(\mathbf{S}_{B} - \mathbf{S}_{C}) \cdot \hat{z}| = \frac{1}{N} \sum_\mathbf{k} |(\mathbf{S}_{\mathbf{k},B} - \mathbf{S}_{\mathbf{k},C}) \cdot \hat{z}| \\
    m_{s,xy} &= |(\mathbf{S}_{B} - \mathbf{S}_{C}) \times \hat{z}| = \frac{1}{N} \sum_\mathbf{k} |(\mathbf{S}_{\mathbf{k},B} - \mathbf{S}_{\mathbf{k},C}) \times \hat{z}|
\end{align}
using the Hartree-Fock decomposition. For details, see Supplementary Material (SM) Sec.~I.

\begin{figure}
    \centering
    \includegraphics[width=\linewidth]{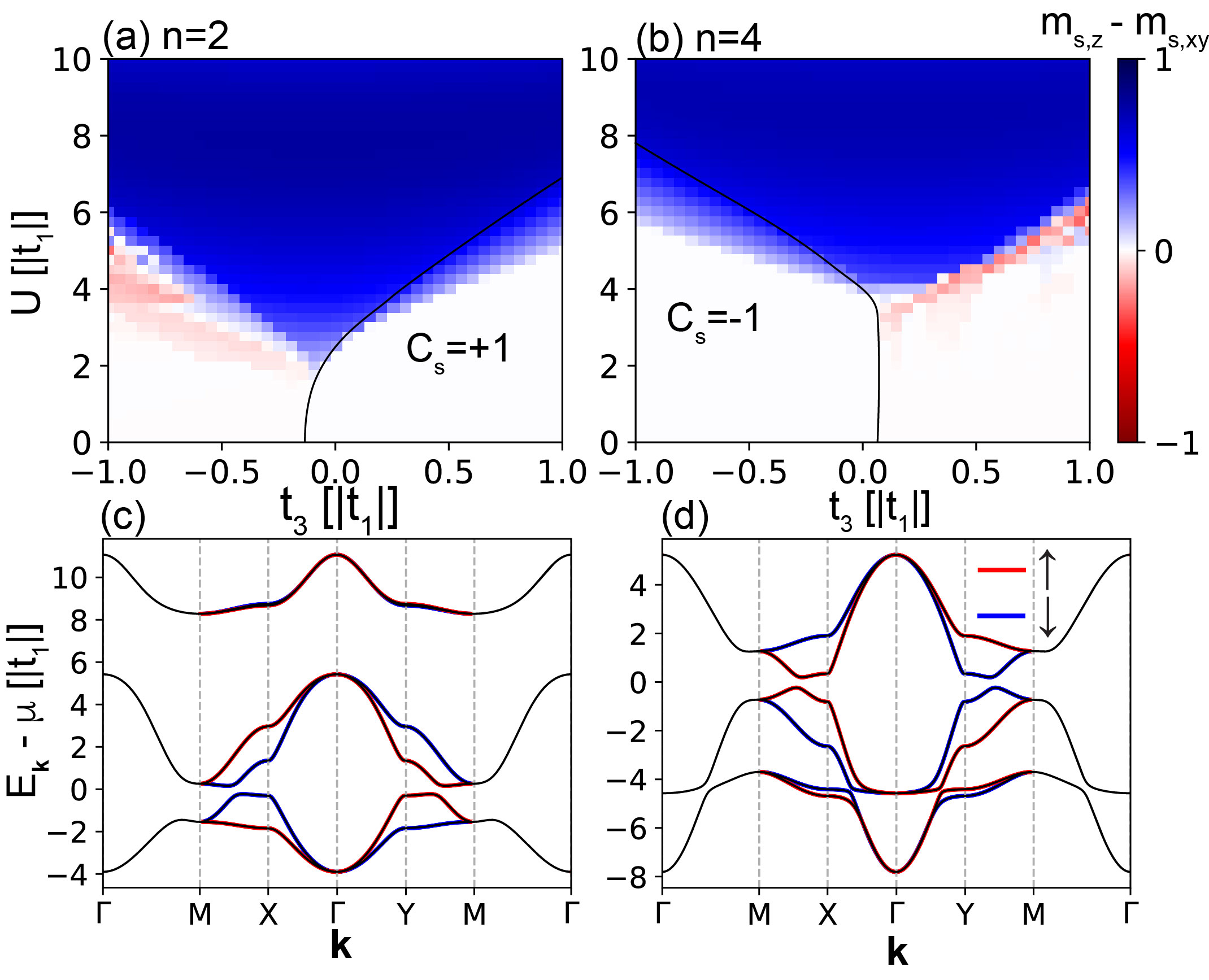}
    \caption{(a,b) Difference in the out-of-plane and in-plane staggered magnetization, (c,d) spin-resolved band structure in the topological altermagnetic insulating phase for (a,c) 2 and (b,d) 4 electrons in the unit cell. Parameters are $t_1=-1$,$t_2=1.5$, $\lambda_I = 0.1$, $\epsilon_A = 10,n=2$ in (a,c) and $\epsilon_A = 0,n=4$ in (b,d), with $t_3,U=0.7,5$ in (c) and $t_3,U = -0.5,5.5$ in (d).}
    \label{fig:Fig1}
\end{figure}

For $n=2$ and $4$ electrons in the unit cell, this gives the phase diagrams in Fig.~\ref{fig:Fig1} (a) and (b), respectively. For small values of $U$, the system is a paramagnet with a quadratic band crossing at the $M$ point for $t_3>(<)0$ and $n=2(4)$ at $\lambda_I = 0$. The Ising SOC then gaps out the crossing, leading to a spin Chern number of $\pm 1$, respectively. Upon increasing $U$, the system acquires antiferromagnetic order on the $B$ and $C$ sublattices, with zero net magnetization, becoming an altermagnet. This splits the quadratic band crossing into two spin-polarized Dirac points along the boundary of the Brillouin zone (see Fig.~\ref{fig:Fig0} (b)). Due to the presence of the Ising SOC, the N\'eel vector is pointing along the $z$-direction, which allows for the Dirac points to be gapped \cite{Antonenko2025}, yielding the band structures in Fig.~\ref{fig:Fig1} (c,d), respectively. 

The topological phase is limited by the value of $U$, where two Dirac points with the same spin merge at the $X$ and $Y$ point, respectively. As a result, for higher $U$ than this critical value, the system becomes a topologically trivial altermagnetic insulator. As the gap therein is opened by suppressing double occupancy on the $B$ and $C$ sites, with the Hartree-Fock calculations yielding charge occupations very close to 1, we designate this as an altermagnetic Mott insulator, similar to Ref.~\cite{Kaushal2025}. This point is further supported by our exact diagonalization results lateron.

On the other hand for $t_3 <(>) 0$ and $n=2(4)$ below the critical value of $U$, respectively, no band inversion with associated Dirac points occurs, leaving the system in a metallic state favoring an in-plane orientation of the N\'eel vector. This phase is terminated by a similar value of $U$ where the topological phase transitions to an altermagnetic Mott insulator and the system once again favors an out-of-plane N\'eel vector.

\begin{figure}
    \centering
    \includegraphics[width=\linewidth]{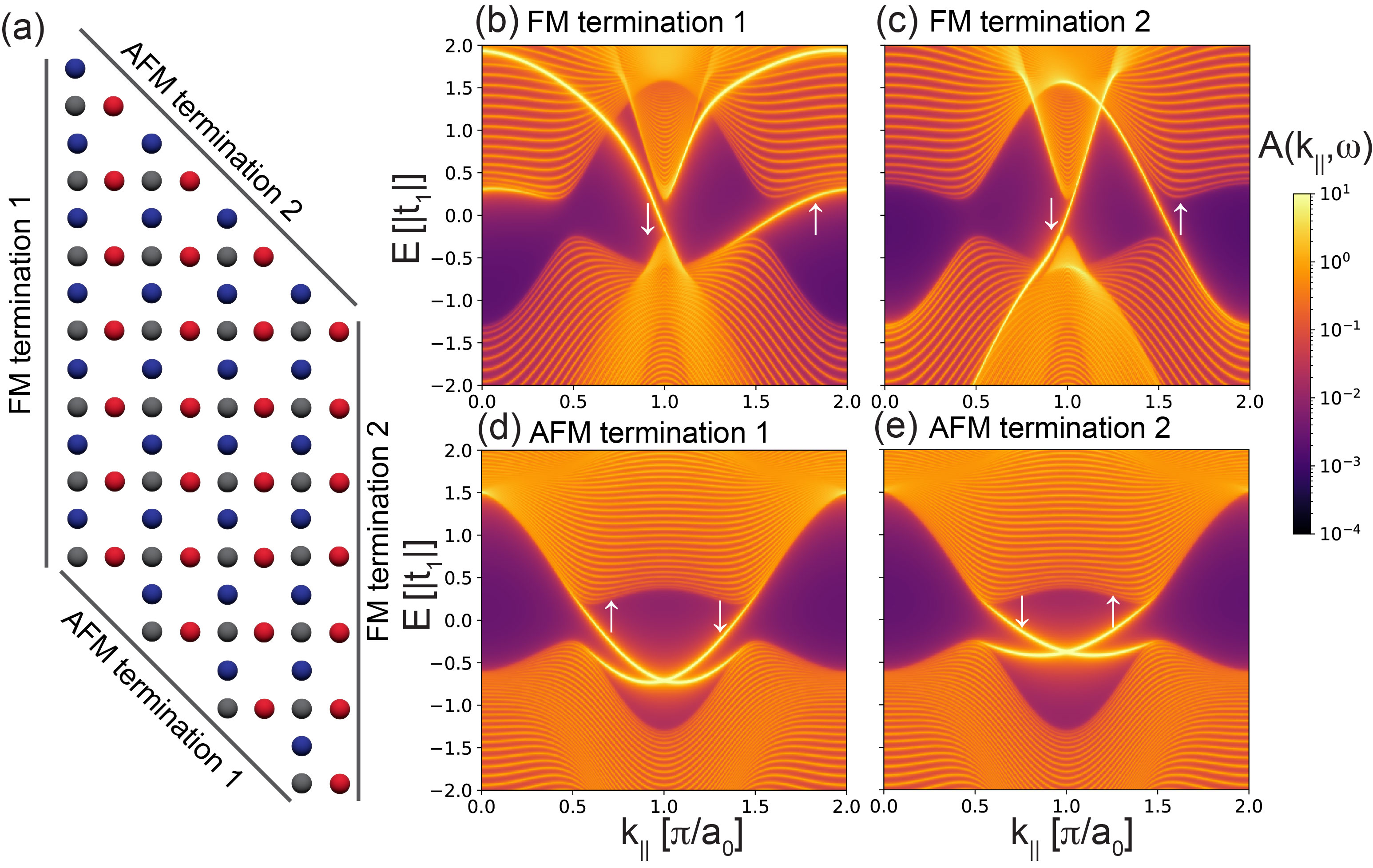}
    \caption{(a) schematic of considered edge types with different magnetic structures and (b) location of the massive Dirac cones in the Brillouin zone, (c-f) spectral function along (c)[(d)] ferromagnetic edges with AC [B] termination, respectively,  
    along (e)[(f)] antiferromagnetic edges with A[BC] termination, respectively. Parameters are $t_1=-1,t_2=1.5,t_3=-0.5,\epsilon_A=0,\lambda_I=0.1,U=5.5,n=4$.}
    \label{fig:Fig2}
\end{figure}

{\it Helical Edge Modes.} We now consider the electronic structure in a long strip, with translation invariance along one direction and open boundary conditions along another, onto which we impose the order parameters obtained from the HF calculations of the translation-invariant system. For the Lieb lattice there are 4 unique types of edge structures to consider: two vertical, edges (equivalent to horizontal edges through the $C_{4z}$ symmetry), with ferromagnetic alignment of the spins, and two diagonal edges, with either antiferromagnetic alignment of the spins or non-magnetic $A$ sublattice sites, (see Fig.~\ref{fig:Fig2}(a)). In the topological altermagnetic phase, each of these edges hosts two localized modes due to $|C_s| = 1$. 

We start by considering the ferromagnetic edges, the left one composed of alternating $A$ and $C$ sites and the right one only by $B$ site. For this system, we present the spin-summed spectral function $A(k_\parallel, \omega)$ at the left and right edge in Fig.~\ref{fig:Fig2}(b),(c). Two edge modes of opposite helicity can be observed in both, resulting in  a clear propagation direction for the electrons. Importantly, the massive Dirac points where the two edge modes originate are completely spin-polarized and separated in momentum for this kind of edge. 

For the antiferromagnetic edges, we present the spectral function $A(k_\parallel, \omega)$  at the left and right edge in Fig.~\ref{fig:Fig2}(d),(e). Here, the two edge modes are symmetric about $k_\parallel = \pi/a_0$, as the diagonal edges lead to the Dirac points being projected onto the same momentum. 

{\it Effective Edge Theory.} A common detrimental effect to topological systems characterized by a spin Chern number is Rashba SOC, as in its presence spin is no longer a good quantum number. To assess the robustness of the helical edge modes against such terms, we formulate a low-energy effective edge theory for the two types of terminations we considered above. 
The topological-trivial transition occurs when two spin-polarized Dirac cones meet at the $X$ and $Y$ points respectively. Assuming without loss of generality that $S<0$ and $t_3<0$, the Dirac cones meeting at the $X$($Y$) point will have spin $\uparrow$($\downarrow$) polarization, respectively. Expanding the Hamiltonian in the respective subspace around these points relative to the Fermi energy then yields
\begin{equation}\label{e6}
    \mathcal{H}_{\uparrow} = \begin{pmatrix}
        \frac{\delta m}{2}& -v_1  (k_x - \pi) - \mathrm{i} v_2 k_y \\
        -v_1 (k_x - \pi) + \mathrm{i} v_2 k_y & - \frac{\delta m}{2}
    \end{pmatrix}
\end{equation}
where $v_1 = 2 \left(t_2 + \frac{t_1^2}{\mu_A}\right)$, $v_2 = 2 \lambda_I$, $\frac{\delta m}{2} = U|S| + 2t_3 - \left(2 - \frac{\pi^2}{2}\right) \frac{t_1^2}{\mu_A}$, such that $\delta m > 0$ in the trivial phase and $\delta m < 0$ in the topological phase. $\mathcal{H}_{\downarrow}$ is obtained by replacing $k_x\to k_y$, $k_y\to -k_x$ and $\delta m\to -\delta m$ in Eq.\ \eqref{e6}.
A Jackiw-Rebbi type solution at a domain wall along the $y$-direction then yields the dispersion relations $E_\uparrow (k_y) = v_2 k_y$, $E_\downarrow (k_y) = - v_1 (k_y - \pi)$.
Hence at low energies, the two helical edge modes are separated in momentum along this kind of edge and similarly for horizontal edges due to the $C_4\mathcal{T}$ symmetry.

On the other hand, for diagonal edges $k_\parallel = k_x + k_y, k_\perp = k_x - k_y$, we obtain 
$E_{\uparrow,\downarrow} = \mp v_{\rm eff}(k_\parallel - \pi)$, 
with $v_{\rm eff}=v_1 v_2/\sqrt{v_1^2 + v_2^2}$, resulting in two helical edge modes crossing at $k_\parallel = \pi$ as observed in Fig.~\ref{fig:Fig2}(e). This is a result of the two spin-polarized Dirac cones being projected onto the same value of $k_\parallel$ (for further details and the corresponding eigenvectors, see SM Sec.~II). 

Now we consider the effect of Rashba SOC between nearest-neighbor sites of strength $\lambda_{R,1}$, which in reciprocal space takes the form
\begin{equation}
\begin{aligned}
    \mathcal{H}_{\lambda_R} &= 2 \lambda_{R,1} \sum_k c_k^\dagger \begin{pmatrix}
        0 & -  s_{k_x/2} \sigma_y & s_{k_y/2} \sigma_x \\
        - s_{k_x/2} \sigma_y & 0 & 0 \\
        s_{k_y/2} \sigma_x & 0 & 0 \\
    \end{pmatrix} c_k  \; ,
\end{aligned}
\label{eq:Rashba}
\end{equation}
where $c^\dagger_k = (c^\dagger_{k,A,\uparrow}, c^\dagger_{k,A,\downarrow}, c^\dagger_{k,B,\uparrow}, c^\dagger_{k,B,\downarrow}, c^\dagger_{k,C,\uparrow}, c^\dagger_{k,C,\downarrow})$ and $s_k = 2\sin(k)$. As this term respects the $C_{4z} \mathcal{T}$ symmetry, it does not change the orientation of the magnetic moments when the N\'eel vector is pointing out-of-plane in the topological phase. 
The inclusion of this term produces fundamentally different effects on the two types of edges discussed above. 

For the diagonal edges, the two opposite-spin Dirac points are projected onto the same value of $k_\parallel$, where expanding around $X$ and $Y$ simultaneously results in a non-zero matrix element of $\mathcal{H}_{\lambda_R}$ between the opposite-spin helical edge modes (for details, see SM Sec. II) , yielding a gap of $\sim 0.288 \lambda_{R,1}$ for the parameters used in Fig.~\ref{fig:Fig2} along diagonal terminations.

For the vertical edges, the separation in momentum of the projected Dirac points prevents the Rashba SOC from producing scattering between the two helical edge modes. Therefore, the helical edge modes along vertical and, by extension, horizontal edges are more robust than their counterparts on diagonal edges, protecting the topological altermagnetic phase.

\begin{figure}
    \centering
    \includegraphics[width=\linewidth]{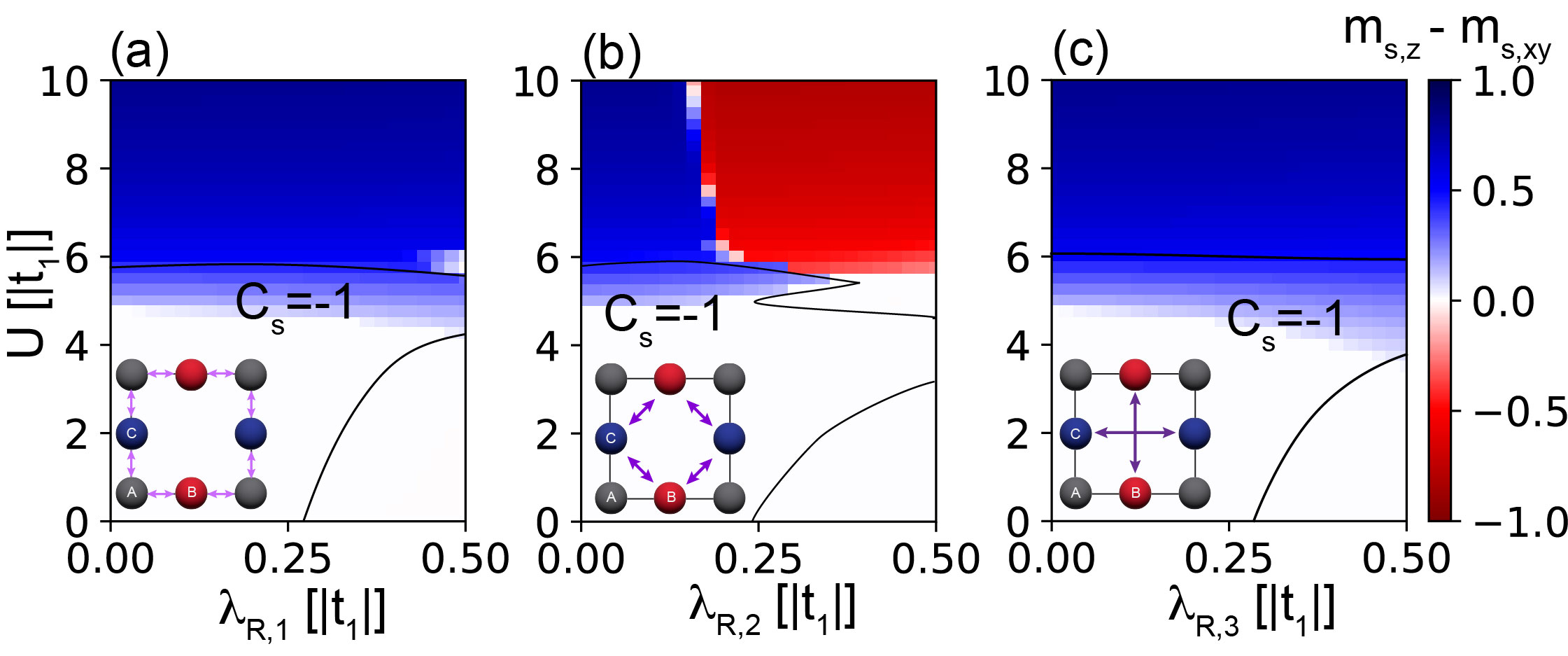}
    \caption{Difference in the out-of-plane and in-plane staggered magnetization as a function of the strength of the Hubbard interaction $U$ and the Rashba SOC $\lambda_{R,m}$ between $m$-th nearest-neighboring sites with (a) $m=1$, (b) $m=2$, and (c) $m=3$. Parameters are $t_1=-1$,$t_2=1.5$, $t_3=-0.5$, $\lambda_I = 0.1$, $\epsilon_A = 10$, $n=4$.}
    \label{fig:Fig3}
\end{figure}

The extent of the topological phase is determined to some degree by the placement of Rashba SOC on different bonds. To assess this, we present in Fig.~\ref{fig:Fig3} the difference of the out-of-plane and in-plane staggered magnetization as a function of the Hubbard interaction and the strength of the Rashba SOC for various placements, as indicated in the insets. Here, we choose other parameters such that there is a substantial region where the system is a topological altermagnetic insulator.

For a Rashba SOC between next-nearest neighbor sites, i.e. between the two different magnetic sublattices, Fig.~\ref{fig:Fig3}(b) shows that the N\'eel vector starts preferring an in-plane orientation above $\lambda_{R,2} \approx 0.2$. Above this value, the altermagnetic topological phase becomes suppressed quickly. 
On the other hand, Fig.~\ref{fig:Fig3}(a) and (c) show that the presence of the Rashba SOC between the non-magnetic and magnetic sublattice and between the same sublattice, respectively, do not affect the orientation of the N\'eel vector, as it retains its preference to point out-of-plane. In fact, the region where the system is both altermagnetic and topological increases with the strength of $\lambda_R$.

\begin{figure}[t]
    \centering
    \includegraphics[width=\linewidth]{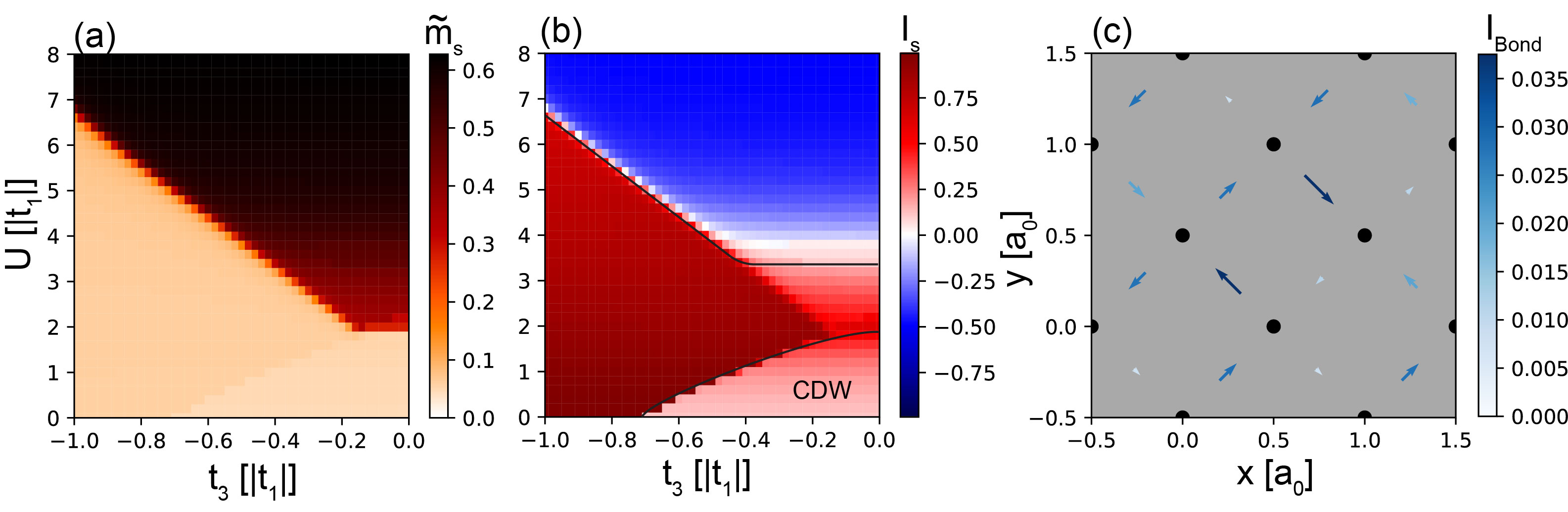}
    \caption{(a) Staggered magnetization, (b) total and (c) bond-wise Ising correlation obtained from exact diagonalization on a $2\times2$ checkerboard cluster at half filling. Parameters are $t_2=1.5$,$\lambda_I = 0.1$ with $U = 3$, $t_3=-0.2$ in (c).}
    \label{fig:Fig4}
\end{figure}

{\it Exact Diagonalization.} In the topological phase obtained in the HF calculations, the $A$ sublattice is either mostly occupied or mostly empty for 4 and 2 electrons per unit cell, respectively. It is therefore possible to omit the $A$ sublattice, reducing the Hamiltonian from Eq.~\eqref{eq:ham_liebHubbard} to a Checkerboard model. Using this effective description we can perform ED studies at filling 2 to confirm the validity of our mean-field results. For a $2\times2$ cluster, we obtained the staggered magnetization  from correlations as $\tilde{m}_s = (N_xN_y)^{-2} \sum_{\alpha \in \{ x,y,z\}} \sum_{i,j} \langle \mathbf{S}_{i,B}^\alpha \left( \mathbf{S}_{j,B}^\alpha - \mathbf{S}_{j,C}^\alpha \right) \rangle, \alpha \in \{x,y,z\}$ , which we show in Fig.~\ref{fig:Fig4}(a) as a function of $U$ and $t_3$. This phase diagram matches our mean-field results qualitatively with large staggered magnetization and suppressed double occupancy for large $U$, supporting an altermagnetic Mott insulating phase, a phase with weak density correlations and small staggered magnetization for large $t_3$, and a charge density wave phase with density correlations and small staggered magnetization for small $U$ and $t_3$ (for further details, see SM Sec. III). Furthermore, similar to our Hartree-Fock results, where these three regions meet, we find a region with intermediate staggered magnetization and small density correlations, matching our expectation for the TAI phase. 
To confirm whether this phase is topological, similar to Ref.~\cite{Sur2018}, we calculate the current correlations between one nearest-neighbor bond $\langle i,j\rangle$ and a reference bond $\langle i_0, j_0 \rangle$ of opposite spin defined as 
\begin{equation} 
I_{\mathrm{Bond},ij} = - \langle (c^\dagger_{i,\uparrow} c_{j,\uparrow} - c^\dagger_{j,\uparrow} c_{i,\uparrow}) (c^\dagger_{i_0,\downarrow} c_{j_0,\downarrow} - c^\dagger_{j_0,\downarrow} c_{i_0,\downarrow}) \rangle.
\end{equation}
We form a staggered sum with the pre-factors of each bond matching the signs of the Ising spin-orbit coupling $I_s = \sum_{\langle i,j \rangle} \nu_{ij} I_{\mathrm{Bond},ij}$, yielding the phase diagram in Fig.~\ref{fig:Fig4}(b), showing positive values of the staggered current correlations in the intermediate phase, as well as the charge density wave and the non-magnetic phase for large $t_3$. From the bond-resolved current correlations, shown in the intermediate phase in Fig.~\ref{fig:Fig4}(c), we see that there is a fixed pattern of the current correlations forming clear loops. This loop pattern is absent in the phase for strong $U$, indicating their qualitative difference.

{\it Discussion.~} In this Article, we showed using Hartree-Fock theory on the Lieb lattice Hubbard model as a minimal model for the anti-CuO$_2$ structure of oxychalcogenides that altermagnetism and topological states can co-exist. For both fillings $n=2$ and $n=4$, a broad region exists, where an altermagnetic semimetal with Dirac cones at the Fermi level becomes a topological altermagnetic insulator due to the presence of Ising spin-orbit coupling. We found that the Ising SOC supports a preferred out-of-plane direction of the N\'eel vector in the region of the parameter space, where the system is insulating. On the other hand, the region of the parameter space that remains metallic prefers an in-plane orientation of the N\'eel vector.

Further, we showed that the magnetic structure along various edges of the system shapes the electronic structure of the helical edge modes induced by the bulk topology. Therein, diagonal non-magnetic and antiferromagnetic edges are much more symmetric than ferromagnetic edges. For ferromagnetic edges, the helical edge modes are separated in momentum at low energies. We showed how this momentum separation of the edge modes emerges from the bulk Hamiltonian in an effective edge theory and that this in fact makes the ferromagnetic edges more robust to inversion-symmetry breaking terms. We confirmed this by analyzing the magnetic order parameters for Rashba SOC along three different bonds in Hartree-Fock theory, showing that the topological altermagnetic region persists to high values of the Rashba spin-orbit coupling.

Finally, we confirmed the stability of the topological phases identified using the mean-field HF approach against quantum fluctuations by employing exact diagonalization techniques. Calculations within the ED on a $2\times 2$ cluster of  staggered magnetization and bond current correlations confirm its altermagnetic character and support the overall topology of the mean-field phase diagram.

{\it Acknowledgments} The authors are indebted to A. Patri for stimulating discussions. This work was supported by the Natural Sciences and Engineering Research Council of Canada (NSERC).

{\bf Data availability}
The data that support the findings of this article are not publicly available. The data are available from the authors upon reasonable request. \\

\end{document}